# Approaches, Challenges and Future Direction of Image Retrieval

Hui Hui Wang, Dzulkifli Mohamad, N.A. Ismail

**Abstract**— This paper attempts to discuss the evolution of the retrieval approaches focusing on development, challenges and future direction of the image retrieval. It highlights both the already addressed and outstanding issues. The explosive growth of image data leads to the need of research and development of Image Retrieval. However, Image retrieval researches are moving from keyword, to low level features and to semantic features. Drive towards semantic features is due to the problem of the keywords which can be very subjective and time consuming while low level features cannot always describe high level concepts in the users' mind. Hence, introducing an interpretation inconsistency between image descriptors and high level semantics that known as the semantic gap. This paper also discusses the semantic gap issues, user query mechanisms as well as common ways used to bridge the gap in image retrieval.

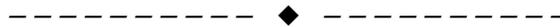

## 1 INTRODUCTION

IMAGE retrieval is the field of study concerned with searching and browsing digital images from database collection. This area of research is very active research since the 1970s [1, 2]. Due to more and more images have been generated in digital form around the world, image retrieval attracts interest among researchers in the fields of image processing, multimedia, digital libraries, remote sensing, astronomy, database applications and other related area. Effective and fast retrieval of digital images has not always been easy, especially when the collections grow into thousands. An effective image retrieval system needs to operate on the collection of images to retrieve the relevant images based on the query image which conforms as closely as possible to human perception.

The purpose of an image database is to store and retrieve an image or image sequences that are relevant to a query. There are a variety of domains such as information retrieval, computer graphics, database management and user behavior which have evolved separately but are interrelated and provide a valuable contribution to this research subject. As more and more visual information is available in digital archives, the need for effective image retrieval has become clear [3,4]. In image retrieval research, researchers are moving from keyword based, to content based then towards semantic based image retrieval and the main problem encountered in the content-based image retrieval research is the semantic gap between the low-level feature representing and high-level semantics in the images

### 1.1 Keyword Based Image Retrieval

In 1970s, the Keyword based Image Retrieval system used keywords as descriptors to index an image. General Framework of keyword based image retrieval is shown in Fig.1.

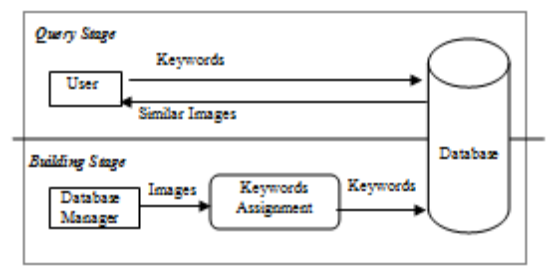

Fig 1. General Framework of Keyword Based Image Retrieval

Before images are being stored in the database, they are examined manually and assigned keywords that are most appropriate to describe their contents. These keywords are stored as part of the attributes associated to the image. During query stage, the image retrieval system will accept from the user one or many keywords which constitute the search criteria. A keyword matching process is then performed to retrieve images associated with the keywords that match the search criteria.

"A picture is worth a thousand words"; this familiar proverb emphasizes that visual information is inherently ambiguous and semantically rich. The content of an image is much richer than what any set of keywords can express, just employing text to describe the content of the image which often causes ambiguity and inadequacy in performing an image database search and query processing. This problem is due to the difficulty in speci-

H.H.Wang, Dzulkifli Mohamad, Nor Azman Ismail are with department of Computer Graphics and Multimedia, Faculty of computer science and information Technology, UTM, Skudai, Malaysia fying exact terms and phrases in describing the content of images as the content of an image is much richer than what any set of keywords can express. Since the textual annotations are based on language, variations in annotation will pose challenges to image retrieval. Comprehensive surveys of early text-based image retrieval methods can be found in [5,6].

## 1.2 Content Based Image Retrieval

In 1980s, Content-based image retrieval (CBIR) then has been used as an alternative to text based image retrieval. IBM was the first, who take an initiative by proposing query-by image content (QBIC). QBIC developed at the IBM Almaden Research Center is an open framework and development technology. Unlike keywords-based system, visual features for contents-based system are extracted from the image itself. CBIR can be categorized based on the type of features used for retrieval which could be either low level or high level features. At early years, low level features include colour (distribution of color intensity across image), texture (Homogeneity of visual patterns), shape (boundaries, or the interiors of objects depicted in the image), spatial relations ( the relationship or arrangement of low level features in space) or combination of above features were used. General Framework of Content based Image Retrieval is shown in Fig.2.

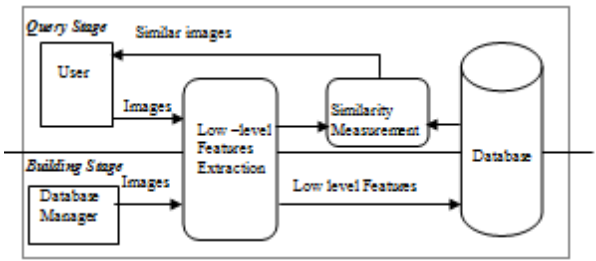

Fig 2. General Framework of Content Based Image Retrieval

All images will undergo the low level feature extraction process before being added to the images database. In feature extraction stage, features such as colour, shape or texture are extracted from the image. User provides a sample image and the similarity measurement engine is responsible in estimating the similarity between the query image and database images and then ranking them according to their similarity to the given query image

The CBIR researches were done in retrieving the image on the basis of their visual content as shown in Table I.

TABLE 1
RESEARCH WORKS ON CONTENT BASED IMAGE RETRIEVAL BASED ON VISUAL CONTENTS

| Low level features | Researches | Approaches |
|---|---|---|
| Color | W. Niblack et al [7] | Histogram and color moments |
| | Chad Carson et al [8] | Region Histogram |
| | J. Sawhney & Hafner [9] | Color Histogram |
| | Stricker & Orengo [10] | Color Moment |
| Shape | MichaelOrtega et al [11] | Fourier Transform |
| | F. Mokhtarian et al [12] | Curvature scale Space |
| | Sougata Mukherjea [13] | Template Matching |
| | Fumikazu Kanehara [14] | Convex parts |
| | Pentland et al [15] | Elastic deformation of templates |
| Texture | J. R. Smith [16] | Wavelet transform |
| | S. Michel[17] | Edge statistic |
| | B. S. Manjunath [18] | Gabor filters |
| | George Tzagkarakis & Panagiotis Tsakalides [19] | Statistical |

Spatial feature is proved useful and effective in grating with other low level features such as colour [20, 21, 22], shape [23, 24] and texture [25] to further increase the confidence in image understanding.

Although there are many sophisticated algorithms to describe color, shape, texture and spatial features approaches, these algorithms do not satisfied and comfort to human perception This is mainly due to the unavailability of low level image features in describing high level concepts in the users' mind such as find an image of a baby is crying loudly. The only way a machine is able to perform automatic extraction is by extracting the low level features that represented by the color, texture, shape and spatial from images with a good degree of efficiency.

## 1.3 Semantic Based Image Retrieval

Neither a single features nor a combination of multiple visual features could fully capture high level concept of images. Besides, due to the performance of Image retrieval based on low level features are not satisfactory, there is a need for the mainstream of the research converges to retrieval based on semantic meaning by trying to extract the cognitive concept of a human to map the low level image features to high level concept (semantic gap). In addition, representing image content with semantic terms allows users to access images through text query which is more intuitive, easier and preferred by the front end users to express their mind compare with using images. For example, users' queries may be 'Find an image of sunset rather than 'find me an image contains red and yellow colors'.General Framework of Semantic based Image Retrieval is shown in Fig. 3.

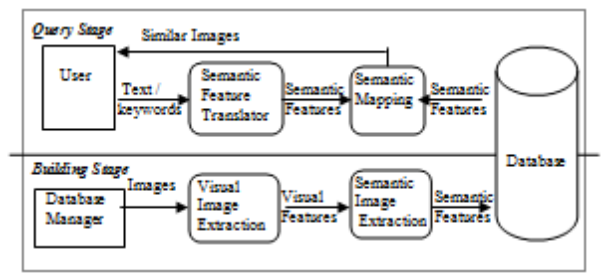

Fig.3 General Framework of Semantic Based Image Retrieval

All images need to go through the visual feature extraction process where it needs to extract the low level features of images to identify meaningful and interesting regions/objects based on the similar characteristics of the visual features. Next, the object/region features will go into Semantic Image Extraction process to get the semantics description of images to be stored in database. Image retrieval can be queried based on high level concept. User can have query based on a set of textual words. Then, it will go into semantic features translator to get the semantic features from the user query. The semantic mapping process is used to find the best concept to describe the segmented region/objects based on the visual features between the query image and database images. This mapping usually will be done through supervise or unsupervised learning tools to associate the low level features with object concept and will be annotated with textual word through image annotation process.

## 2 SEMANTIC GAP

Bridging the semantic gap for image retrieval is a very challenging problem yet to be solved [26,27]. The semantic gap is described as

*...the lack of coincidence between the information that one can extract from the visual data and the interpretation that the same data have for a user in a given situation. [30]*

Describing images in semantic terms is an important and challenging task that needed to carry out to fulfill human satisfaction besides to have more intelligent image retrieval system.

Human beings are able to interpret images at different levels, both in low level features (colour, shape, texture and object detection) and high level semantics (abstract objects, an event). However, a machine is only able to interpret images based on low level image features. Besides, users prefer to articulate high-level queries [28,29], but CBIR systems index images using low-level features. Hence, introducing an interpretation inconsistency between image descriptors and high-level semantics that is known as the semantic gap [3,29]. The semantic gap is the lack of correlation between the semantic categories that a user requires and the low-level features that CBIR systems offer. The semantic gap between the low-level visual features (color, shape, texture, etc.) and semantic concepts identified by the user remains a major problem in content based image retrieval [27].

Semantic content representation has been identified as an important issue to bridge the semantic gap in visual information access. It has been addressed as a good description and representation of an image, it able to capture meaningful contents of the image. Besides, users preferred to express their information needs at the semantic level instead of the level of preliminary image features. Moreover textual queries usually provide more accurate description of users' information needs.

## 2.1 Bridging the Semantic Gap: Mapping From low level features to High level concept

A major challenge Image Retrieval is to achieve meaningful mappings that minimize the semantic gap between the high-level semantic concepts and the low-level visual features in images. High-level concepts, however, are not extracted directly from visual contents, but they represent the relatively more important meanings of objects and scenes in the images that are perceived by human beings. These conceptual aspects are more closely related to users' preferences and subjectivity. Although the semantic concepts are usually not directly related to the visual image features (color, texture, shape and spatial relation), these attributes capture information about the semantic meaning. So, in order to bridge the feature-to-concept gap, effective extraction and selection of low level features are needed for correlating with high level semantic description for image interpretation and to achieve more intelligent and user friendly retrieval. Other than conventional low-level visual features, new features should be learned, which are more representative to describe the semantic meaning of concepts [31]

The attempt to overcome the gap between high-level semantics and low-level features by representing images at the object level is needed [32]. The semantic extraction and representation of images process is shown in Fig.4

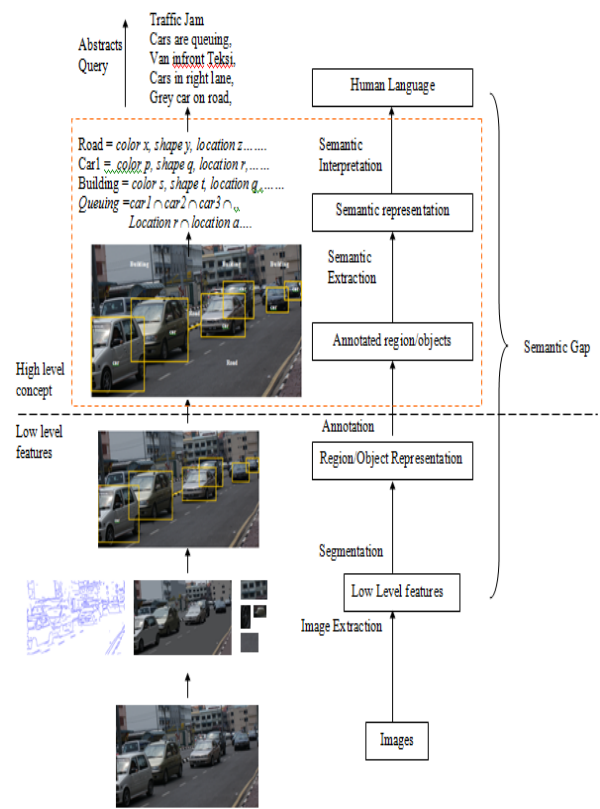

Fig.4 Bridging the gap: the semantic extraction and representation of images

Semantic representation of images can be done through the process as shown in Fig 4. Firstly, the image extraction process will get the low level features of images either by color, shape, textures and spatial. Next, these low level features can be clustered or segmented based on the similar characteristics of the visual features to form some regions representation and next to form objects identification and representation in the images. The regions/objects representation will be annotated with keyword by image annotation process. This annotation process can be done either manually, semi-automatically or automatically. The image then will be represented using semantics and image retrieval can be queried based on high level concept. Semantic content representation has been identified as an important issue to bridge the semantic gap in visual information access. It has been addressed as a good description and representation of an image and is able to capture meaningful contents of the image.

Table II shows the common techniques used in bridging the semantic gap in image retrieval or in other words, mapping from low level to high level concepts.

TABLE II
RESEARCH WORKS ON BRIDGING THE SEMANTIC GAP IN IMAGE RETRIEVAL

| Techniques | | Researchers | Approaches |
|---|---|---|---|
| Annotation | Manual | Wikipedia image collection[33], google image labeler [34] | User annotated –whole image |
| | | Inotes [35] | User annotated –region /object |
| | | Facebook [36] | User annotated –region /object |
| | Semi-Automatic | Bradshaw [37] | Bayer probability |
| | | Ghoshal *et al* [38] | Co-Occurrence model |
| | | Ilaria Bartolini and Paolo Ciaccia[39] | Graph based link |
| | Automatic | Huang *et al* [40] | Decision Trees |
| | | Feng and Chua [41] | Bootstrapping |
| | | Gao et al. [42] | Latent Semantic Analysis |
| | | Mori et al. [43] | hidden Markov model |
| Relevance feedback | | Yang et al. [44] | Semantic feedback mechanism |
| | | Rege et al. [45] | multi-user relevance feedback(user-centered semantic hierarchy) |
| | | Lin et al.[46] | semantic manifold |
| | | Janghyun Yoon and Nikil Jayant [47] | multi-modal model feedback |
| Object Ontology | | P.L.Standchey et al [48] | Color representation ontology |
| | | V. Mezaris [49] | High level concept ontology |
| | | Huan Wang [50] | multi-modality ontology |
| Semantic template | | S.-F. Chang [51] | Semantic visual template |

## 3 USER QUERY

Query mechanisms play a vital role in bridging the semantic gap between users and retrieval systems [52]. The user query is used to express the user's information need to retrieve images in collection of database that conform to human perception. The quality of queries submitted to information retrieval (IR) systems directly affects the quality of search results generated [53] According to Ref [54], to define a semantic meaning and representation of the input query that can precisely understand and distinguish the intent of the input query as well as the domain coverage are the major challenges. It is difficult and often requires many human efforts to meet all these challenges by the statistical machine learning approaches.

In Ref. [55], Eakins mentioned three levels of queries in CBIR.

*Level* 1: Retrieval by primitive features such as color, texture, shape or the spatial location of image elements. Typical query is query by example, 'find pictures like this (given sample image)'.

*Level* 2: Retrieval of objects of given type identified by derived features, with some degree of logical inference. For example, 'find a picture of a bus'.

*Level* 3: Retrieval by abstract attributes, involving a significant amount of high-level reasoning about the purpose of the objects or scenes depicted. This includes retrieval of named events, of pictures with emotional or religious significance. Query example, 'find pictures of a happy and cheerful girl'.

Levels 2 and 3 together are referred to as semantic image retrieval, and the gap between Levels 1 and 2 as the semantic gap [55].

The user query can be classified into 2 main groups.

### 3.1 Query by Visual Example

Querying by visual example [56, 57, 58] is a paradigm, particularly suited to express perceptual aspects of low/intermediate features of visual content [29]. Visual content refer to color, shape and texture features of images. However, users will not always have an example image on hand and also it's hard to find a suitable example to describe what is in user mind. Efforts have been made to extend the query by visual example to query by region selection [51, 59], sketch [60, 61, 62]. Users are allowed to select their "region of interest' in the image or draw their desired content. Although promising progresses have been made in image retrieval techniques based on visual features, formulating a query such as submitting an example image or a sketch is sometimes not convenient for users besides unable to fully describe user needs.

### 3.2 Query by Texts

User usually prefers using keywords to indicate what they want. [63, 64]. Textual queries usually provide more accurate description of users' information needs as it allow users to express their information needs at the semantic level and high level abstractions instead of limited to the level of preliminary image features. However, the textual words need to be translated automatically to semantic meaning and representation that are matched in the images semantic representation in database in order to have fully and precisely understand the user input.

## 4 SUMMARY OF IMAGE RESEARCH

The summary of image retrieval research is shown in Fig.5.

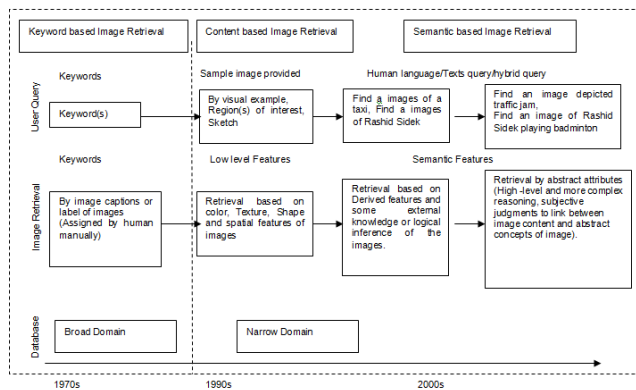

Fig.5 Summary of image retrieval research

In image retrieval research, researchers are moving from keyword based, content based then towards semantic based image retrieval. Human language or high level concept user query is used now compared with 1970s and 1980s where keyword that based on image caption and query by example image were used. The features of image are moving from low level features towards semantic image features that are closest to human perception. However for the domain of dataset, researchers are moving from broad domain to narrow domain due to narrow domain can used for knowledge representation and images share and having limited concept.

## 5 FUTURE DIRECTION: TOWARD INTELLIGENT IMAGE RETRIEVAL

Even though the image retrieval is moving towards semantic concept however, much initial research in semantic image retrieval are focusing just on the simple semantic retrieval such as retrieval of objects of a given type but pays little attention on the retrieval by abstract attributes, involving a significant amount of high-level reasoning about the meaning and purpose of the objects or scenes depicted. In other words, the retrieval by abstract attributes was still not satisfied to human perception. Moreover, the retrieval involved human interference and is time consuming besides inconsistency. There is a need to further increase the confidence in image understanding and to effectively retrieve similar images that are conform to human perception and without human interference.

The semantic gap is harder to overcome in broad domains database rather than narrow domains because images in broad domains can be described using various concepts that are very challenging to detect [65] Besides, broad domains contains images of various scenes, themes, objects and people gathered from the web or from a stock image collection with high visual granularity [66] while narrow domains describe little or limit concepts that are much more easy to detect since the scenes, themes, objects and people having low visual granularity. The researchers are moving to reduce the semantic gap in broad domains.

Researchers are moving towards to intelligent image retrieval that are also supports more abstract in concept by understanding the image content in terms of high level concepts, which is closely related to the problem of computer vision and object recognition besides more intelligent system. The domain should be not specific but broad where all the extracted semantic features are applicable for any kind of images collection. There are still some spaces which need to be improved besides the challenges that are associated with mapping low level to high level concepts such as friendly user interface as well as an efficient indexing tool are needed for contributing to successful Image Retrieval.

## 6 CONCLUSION

As conclusion, this paper provides a study of image retrieval work towards narrowing down the 'semantic gap'. Recent works are mostly lack of semantic features extraction and user behavior consideration. Therefore, there is a need of image retrieval system that is capable to interpret the user query and automatically extract the semantic feature that can make the retrieval more efficient and accurate.

Image retrieval researches are moving from keyword toward semantic based image retrieval. However, existing image retrieval researches are still lack of meaningful semantic image description and user behavior consideration. For user query, textual queries are usually can provide more accurate description of users' information needs. Therefore, there is a need to provide maximum support towards bridging the semantic gap between low level visual features and high level concepts for better image understanding between human and machine and also contribute to have more intelligent, user friendly besides accuracy and efficiency image retrieval.

**H.H.Wang** received the Bachelor of Information Technology and the Master of science from Universiti Malaysia Sarawak (UNIMAS) in 2001 and 2003 respectively. Her master was on Content Based Image Retrieval. Her intelligent image finder project won the gold metal, Swiss government special award for information solution from 34[th] Geneva international invention exposition, 2006. She served as a lecturer at UNIMAS and she is now a Ph.D candidate in Department of Computer Graphics, Faculty of Computer Science and Information Technology, Universiti Teknologi Malaysia (UTM). Her main research interests are in image retrieval, computer vision and Pattern recognition.

**Dzulkifli Mohamad** completed his degree in Computer Science at Universiti Kebangsaan Malaysia and Advanced Diploma in Computer Science at Glasgow University, Scotland. He continued his master's degree and PhD in Computer Science both at Universiti Teknologi Malaysia. .He is currently a professor in the Faculty of computer science and Information Technology at Universiti Teknologi Malaysia. His research interests include areas such as Artificial Intelligence and Identification System

**N.A. Ismail** received his B.Sc. from Universiti Teknologi Malaysia, UTM, Master of Information Technology (MIT) from National University of Malaysia, and Ph.D. in the field of Human Computer Interaction (HCI) from Loughborough University. He has been a lecturer at Computer Graphics and Multimedia Department, Universiti Teknologi Malaysia for about thirteen years and currently, he is Research Coordinator of the Department. He has made various contributions to the field of Human Computer Interaction (HCI) including research, practice, and education.